\documentclass[11pt]{article}
\usepackage{amsfonts}
\usepackage{latexsym,amssymb,amsfonts}
\usepackage{mathrsfs}
\usepackage{graphicx}
\usepackage{psfrag}
\usepackage{amsmath, amsbsy}
\usepackage{amsopn, amstext}
\usepackage{amsmath,multicol,amsopn}
\usepackage{latexsym,amssymb}

\def\sqr#1#2{{\vcenter{\vbox{\hrule height.#2pt
              \hbox{\vrule width.#2pt height#1pt \kern#1pt \vrule width.#2pt}
              \hrule height.#2pt}}}}
\def\signed #1{{\unskip\nobreak\hfil\penalty50
              \hskip2em\hbox{}\nobreak\hfil#1
              \parfillskip=0pt \finalhyphendemerits=0 \par}}
\def\endpf{\signed {$\sqr69$}}
\def\3n{\negthinspace \negthinspace \negthinspace }
\def\2n{\negthinspace \negthinspace }
\def\1n{\negthinspace }

\def\see{{\it see} }
\def\ds{\displaystyle}

\def\={\buildrel \triangle \over =}

\def\l{\lambda}

 \def\n{\nabla}

\def\t{\times}

\def\o{\omega}

\def\ns{\noalign{\ss} }

\def\pa{\partial}
\def\G{\Gamma}
\def\D{\Delta}

\def\Si{\Sigma}

\def\O{\Omega}

\def\cA{{\cal A}}

\def\cF{{\cal F}}

\def\no{\noindent}

\def\ms{\medskip}
\def\bs{\bigskip}
\def\q{\quad}
\def\qq{\qquad}

\def\max{\mathop{\rm max}}
\def\min{\mathop{\rm min}}

\def\pa{\partial}

\def\cd{\cdot}

\def\div{\hbox{\rm div$\,$}}

\def\|{\Big |}
\def\({\Big (}
\def\){\Big )}
\def\[{\Big[}
\def\]{\Big]}
\def\be{\begin{equation}}
\def\bel{\begin{equation}\label}
\def\ee{\end{equation}}
\def\bt{\begin{theorem}}
\def\bcd{\begin{condition}}
\def\ecd{\end{condition}}
\def\et{\end{theorem}}
\def\bc{\begin{corollary}}
\def\ec{\end{corollary}}
\def\bde{\begin{definition}}
\def\ede{\end{definition}}
\def\bl{\begin{lemma}}
\def\el{\end{lemma}}
\def\bp{\begin{proposition}}
\def\ep{\end{proposition}}
\def\br{\begin{remark}}
\def\er{\end{remark}}
\def\ba{\begin{array}}
\def\ea{\end{array}}
\def\ed{\end{document}}
\def\ns{\noalign{\ms}}
\def\ds{\displaystyle}

\def\square#1{\vbox{\hrule\hbox{\vrule height#1%
     \kern#1\vrule}\hrule}}
\def\rectangle#1#2{\vbox{\hrule\hbox{\vrule height#1%
     \kern#2\vrule}\hrule}}

\font\tenbb=msbm10 \font\sevenbb=msbm7
\font\fivebb=msbm5

\newfam\bbfam
\scriptscriptfont\bbfam=\fivebb
\textfont\bbfam=\tenbb
\scriptfont\bbfam=\sevenbb

\newtheorem{lemma}{Lemma}[section]
\newtheorem{remark}{Remark}[section]

\newtheorem{theorem}{Theorem}[section]
\newtheorem{corollary}{Corollary}[section]

\newtheorem{definition}{Definition}[section]
\newtheorem{proposition}{Proposition}[section]
\newtheorem{condition}{Condition}[section]

\makeatletter
   
   \@addtoreset{equation}{section}
\makeatother

\begin{document}
\title{\bf Global Uniqueness for an Inverse Stochastic Hyperbolic Problem with Three Unknowns\thanks{This work is partially supported by the NSF of China under grants
 10831007 and 60974035 and the Grant MTM2011-29306 of the MICINN, Spain.
\ms}}

\author{Qi L\"u \thanks{School of
Mathematical Sciences, University of Electronic Science and
Technology of China, Chengdu 610054, China; and BCAM-Basque Center
for Applied Mathematics, Alameda de Mazarredo, 14, E-48009 Bilbao,
Basque Country, Spain. {\small\it E-mail:} {\small\tt
luqi59@163.com}.\ms} \q and \q Xu Zhang\thanks{School of
Mathematics, Sichuan University, Chengdu 610064, China. {\small\it
E-mail:} {\small\tt zhang$\_$xu@scu.edu.cn}.}}

\date{}

\maketitle

\begin{abstract}\no
This paper is addressed to an inverse stochastic hyperbolic problem
with three unknowns, i.e., a random force intensity, an initial
displacement and an initial velocity. The global uniqueness for this
inverse problem is proved by means of a new global Carleman estimate
for the stochastic hyperbolic equation. It is found that both the
formulation of stochastic inverse problems and the tools to solve
them differ considerably from their deterministic counterpart.
\end{abstract}

\bs

\no{\bf 2000 Mathematics Subject
Classification}.  Primary 60H15; Secondary
65M32.

\bs

\no{\bf Key Words}. Global uniqueness, inverse stochastic hyperbolic
problem,  Carleman estimate, three unknowns.

 \ms


\section{Introduction }

Let $T > 0$, and let $G \subset \mathbb{R}^{n}$ ($n \in \mathbb{N}$)
be a given bounded domain with a $C^{2}$ boundary $\G$. Put $Q \=
(0,T) \t G$ and $\Si \= (0,T) \t \G$. Fix a complete filtered
probability space $(\O, {\cal F}, \{{\cal F}_t\}_{t \geq 0}, P)$, on
which a  one dimensional standard Brownian motion $\{ B(t) \}_{t\geq
0}$ is defined. For any Banach space $H$, denote by $L^{2}_{\cal
F}(0,T;H)$ the Banach space consisting of all $H$-valued $\{ {\cal
F}_t \}_{t\geq 0}$-adapted processes $X(\cdot)$ such that
$\mathbb{E}(|X(\cdot)|^2_{L^2(0,T;H)}) < \infty$, by
$L^{\infty}_{\cal F}(0,T;H)$ the Banach space consisting of all
$H$-valued $\{ {\cal F}_t \}_{t\geq 0}$-adapted bounded processes,
and by $L^{2}_{\cal F}(\O;C([0,T];H))$ the Banach space consisting
of all $H$-valued $\{ {\cal F}_t \}_{t\geq 0}$-adapted continuous
processes $X(\cdot)$ such that
$\mathbb{E}(|X(\cdot)|^2_{C([0,T];H)}) < \infty$. All of these
spaces are endowed with the canonical norm (Similarly, one can
define $L^{2}_{\cal F}(\O;C^{k}([0,T];H))$ for any positive integer
$k$).

Throughout this paper, we assume that the functions  $ b^{ij} \in
C^1(\overline G)$ ($i,j=1,2,\cdots,n$) satisfy  $b^{ij} = b^{ji}$
and, for some constant $s_0
> 0$,
\begin{equation} \sum_{i,j=1}^nb^{ij}\xi^{i}\xi^{j} \geq s_0 |\xi|^2,
\,\,\,\,\,\,\,\,\,\, \forall\, (x,\xi)\in G \t \mathbb{R}^{n}.
\label{bij}
\end{equation}
Consider the following stochastic hyperbolic equation:
\begin{eqnarray}{\label{system1}}
\left\{
\begin{array}{lll}\ds
\ds dz_{t} - \sum_{i,j=1}^n(b^{ij}z_{x_i})_{x_j}dt = \left(b_1 z_t +
b_2\cd\nabla z + b_3 z \right)dt + (b_4z + g)dB(t) & {\mbox { in }}
Q,
 \\
\ns\ds  z = 0 & \mbox{ on } \Si, \\
\ns\ds  z(0) = z_0,\ z_{t}(0) = z_1 & \mbox{ in } G.
\end{array}
\right.
\end{eqnarray}
Here, $z_t=\frac{\pa z}{\pa t}$, $z_{x_i}=\frac{\pa z}{\pa x_i}$,
and $b_i$ $(1 \leq i \leq 4)$ are some suitable known functions to
be given later; while $(z_0, z_1) \in L^2(\O,{\cal F}_0, P; H^1_0(G)
\t L^2(G))$ and  $g\in L^2_{\cal
 F}(0,T;L^2(G))$ are unknown. Physically, $g$ stands for the
intensity of a random force of the white noise type. Put \be H_{T}
\= L_{\cal F}^2 (\O; C([0,T];H_{0}^1(G)))\bigcap L_{\cal F}^2 (\O;
C^{1}([0,T];L^2(G))). \label{HT} \ee It is clear that $H_T$ is a
Banach space with the canonical norm. Under suitable assumptions
(the assumptions in this paper are enough), for any given $(z_0,
z_1)$ and $g$, one can show that the equation (\ref{system1}) admits
one and only one solution $z=z(z_0,z_1,g)(t,x,\omega)\in H_T$ (see
\cite{Zhangxu3}). We will also denote by $z(z_0,z_1,g)$ or
$z(z_0,z_1,g)(t)$ the solution of (\ref{system1}).

In this article, the random force $\int_0^tgdB$ is assumed to cause
the random vibration starting from some initial state $(z_0, z_1)$.
Roughly speaking, our aim is to determine the unknown random force
intensity $g$ and the unknown initial displacement  $z_0$ and
initial velocity $ z_1$ from the (partial) boundary observation
$\left.\frac{\pa z}{\pa \nu}\right|_{(0,T)\times \G_0}$ and the
measurement on the terminal displacement $z(T)$, where $\nu =\nu(x)$
denotes the unit outer normal vector of $G$ at $x\in\G$, and $\G_0$
is a suitable open subset (to be specified later) of $\G$. More
precisely, we are concerned with the following global uniqueness
problem: {\it Do $\left.\frac{\pa z}{\pa
\nu}(z_0,z_1,g)\right|_{(0,T)\times \G_0}=0$ and $z(z_0,z_1,g)(T)=0$
in $G$, $P$-a.s. imply that $g=0$ in $Q$ and $z_0=z_1=0$ in $G$,
$P$-a.s.?}

In the deterministic setting, there exist numerous literatures
addressing the inverse problem of PDEs (See \cite{LRS,Isakov1} and
the rich references cited therein). A typical deterministic inverse
problem close to the above one is as follows: Fix suitable known
functions $a(\cd,\cd)$ and $f_1(\cd,\cd)$ satisfying
$\ds\min_{(t,x)\in Q}|f_1(t,x)|>0$, and consider the following
hyperbolic equation:
 \begin{eqnarray}{\label{sy2}}
\left\{
\begin{array}{lll}\ds
\ds z_{tt} - \Delta z = a(t,x) z +f_1(t,x)f_2(x) & {\mbox { in }} Q,
 \\
\ns\ds  z = 0 & \mbox{ on } \Si, \\
\ns\ds  z(0) = 0,\ z_{t}(0) = z_1 & \mbox{ in } G.
\end{array}
\right.
\end{eqnarray}
In (\ref{sy2}), both $z_1$ and $f_2$ are unknown and one expects to
determine them through the boundary observation $\left.\frac{\pa
z}{\pa \nu}\right|_{(0,T)\times \G_0}$. As shown in \cite{YZ}, by
assuming suitable regularity on functions $a(\cd,\cd)$,
$f_i(\cd,\cd)$ ($i=1,2$) and $z_1(\cd)$, and using the following
transformation
  \begin{eqnarray}{\label{sy3}}
y=y(t,x)=\frac{d}{dt}\left(\frac{z(t,x)}{f_1(t,x)}\right),
 \end{eqnarray}
this inverse problem can be reduced to deriving the so-called
observability for the following wave equation with memory
 $$
\left\{
\begin{array}{lll}\ds
\ds y_{tt} - \Delta y = a_1 y_t + a_2\cd\nabla y + a_3
y\\
\ns\ds  \qq\qq\q+\int_0^t\left[c_1(t,s,x)y(s,x)+c_2(t,s,x)\cd \nabla
y(s,x)\right]ds & {\mbox { in }} Q,
 \\
\ns\ds  y = 0 & \mbox{ on } \Si, \\
\ns\ds  y(0,x) = \frac{z_1(x)}{f_1(0,x)},\ \ y_{t}(0,x) =
f_2(x)-\frac{2\pa_tf_1(0,x)}{|f_1(0,x)|^2} z_1(x) & \mbox{ in } G,
\end{array}
\right.
 $$
where $a_i(\cd,\cd)$ ($i=1,2,3$) and $c_i(\cd,\cd,\cd)$ ($i=1,2$)
are suitable functions. Concerning this problem, if $z_1$ is known
and both functions $a(\cd,\cd)$ and $f_1(\cd,\cd)$ are independent
of the space variable $x$, i.e., there is only one unknown in
(\ref{sy2}), then the corresponding inverse problem is now
well-understood (e.g. \cite{PY,YZ01} and the references therein).
The main tool in the later case is to use the Duhamel principle,
instead of the transform (\ref{sy3}), to reduce the problem to the
observability estimate for some wave equation (without memory).

Stochastic partial differential equations (PDEs for short) are used
to describe a lot of random phenomena appeared in physics,
chemistry, biology, control theory and so on. In many situations,
stochastic PDEs are more realistic mathematical models than the
deterministic ones. Nevertheless, compared to the deterministic
setting, there exist a very limited works addressing inverse
problems for stochastic PDEs. In this respect, we mention
\cite{BCLZ} for a study of an inverse medium scattering problem for
the random Helmholtz equation. Also, we refer to \cite{CT,IK} for
several results on the estimation problems of some random and
stochastic PDEs when the noise intensity tends to zero. To the best
of our knowledge, there is no paper considering the inverse problem
for stochastic hyperbolic equations.

One may meet substantially new difficulties in the study of some
inverse problems for stochastic PDEs. For instance, unlike the
deterministic PDEs, the solution of a stochastic PDE is usually
non-differentiable with respect to the variable with noise (say, the
time variable considered in this paper). Also, the usual compactness
embedding result does not remain true for the solution spaces
related to stochastic PDEs. These new phenomenons lead that some
effective methods for solving inverse problems for deterministic
PDEs (see \cite{PY} for example) cannot be used to solve the
corresponding inverse problems in the stochastic setting.
Especially, one can see that none of the methods for solving the
above inverse problem for the equation (\ref{sy2}) can be easily
adopted to solve our inverse problem for the stochastic hyperbolic
equation (\ref{system1}), even if $g$ is assumed to be of the form
 \bel{po1}
 g(t,x,\o)=g_1(t,\o)g_2(x),\qq \forall\; (t,x,\o)\in Q\times \O,
 \ee
with a known nonzero stochastic process $g_1(\cd,\cd)\in L^2_{\cal
 F}(0,T)$ and an unknown deterministic function $g_2(\cd)\in L^2(G)$.  For these reasons, it is necessary to develop new
methodology and technique for treating inverse problems for
stochastic PDEs.

In this paper, we will use a global Carleman estimate to solve the
above formulated inverse problem for the equation \eqref{system1}.
As far as we know, \cite{Zhangxu3} is the only published reference
addressing the Carleman estimate for stochastic hyperbolic
equations. In \cite{Zhangxu3}, under suitable assumptions, the
following estimate was proved for the solution $z$ of
\eqref{system1}:
\begin{equation} \label{obser esti22}
\begin{array}{ll}\ds
|(z(T),z_t(T))|_{L^2(\O,{\cal F}_T, P; H_0^1(G)\t L^2(G))}
 \leq C \left[\left|\frac{\pa z}{\pa \nu}\right|_{L^2_{\cal
 F}(0,T;L^2(\G_0))} + |g|_{L^2_{\cal
 F}(0,T;L^2(G))}\right].
 \end{array}
 \end{equation}
(Here and henceforth,  $C$ is a generic positive constant, depending
only on $T$, $G$, $\G_0$ and $s_0$, which may be different from one
place to another). Noting however that the (random) source $g$
appears in the right hand side of \eqref{obser esti22}, and
therefore, the estimate obtained in \cite{Zhangxu3} does not apply
to the inverse problem considered in this work. In order to solve
our stochastic inverse problem, we have to establish a new Carleman
estimate for \eqref{system1} so that the source term $g$ can be
bounded above by the observed data. Hence, we need to avoid
employing the usual {\it energy estimate} because, when applying
this sort of estimate to \eqref{system1}, the source term $g$ would
appear as a bad term. Meanwhile, since  we are also expected to
identity the initial data, we need to bound above the initial data
by the observed data, too. Because of this, we need to obtain the
estimate on the initial data and source term in the Carleman
inequality simultaneously. Therefore we cannot use the usual
``Carleman estimate" $+$ ``energy estimate" method (which works well
for the deterministic wave equation, see \cite{Fu-Yong-Zhang1}) to
derive the desired estimates. This is the main difficulty that we
need to overcome in this paper.

The rest of this paper is organized as follows. In Section 2, we
state the main result of this paper. Some preliminary results are
collected in Section 3. Finally, Section 4 is addressed to proving
the main result.


\section{Statement of the main result}


To begin with, we introduce the following
conditions:
\begin{condition}
\label{condition of d} There exists a positive
function $d(\cdot) \in
C^2(\overline{G})$ satisfying the following:\\

\ms

1) For some constant $\mu_0 > 0$, it holds that
\begin{equation}\label{d1}
\begin{array}
{ll} \ds \sum_{i,j=1}^n\Big\{
\sum_{i',j'=1}^n\Big[
2b^{ij'}(b^{i'j}d_{x_{i'}})_{x_{j'}} -
b^{ij}_{x_{j'}}b^{i'j'}d_{x_{i'}} \Big]
\Big\}\xi^{i}\xi^{j} & \!\!\ds\geq \mu_0
\sum_{i,j=1}^nb^{ij}\xi^{i}\xi^{j}, \\
\ns & \ds \forall (x,\xi^{1},\cdots,\xi^{n}) \in \overline{G}  \t
\mathbb{R}^n;
\end{array}
\end{equation}

\ms

2) There is no critical point of $d(\cdot)$ in $\overline{G}$, i.e.,
\be\label{d2}\min_{x\in \overline{G} }|\nabla d(x)| > 0. \ee
\end{condition}

\begin{remark}
If $(b^{ij})_{1\leq i,j\leq n}$ is the identity matrix, then, by
taking $d(x)=|x-x_0|^2$ with $x_0\notin \overline G$, one sees that
Condition \ref{condition of d} is satisfied. Condition
\ref{condition of d} was introduced in \cite{Fu-Yong-Zhang1} to show
the observability estimate for hyperbolic equations. We refer to
\cite{Fu-Yong-Zhang1} for more explanation on Condition
\ref{condition of d} and illustrative examples.
\end{remark}

It is easy to see that if $d(\cdot) \in C^2(\overline{G})$ satisfies
Condition \ref{condition of d}, then for any given constants $a \geq
1$ and $b \in \mathbb{R}$, the function $\tilde{d} = ad + b$ still
satisfies Condition  \ref{condition of d}  with $\mu_0$ replaced by
$a\mu_0$. Therefore  we may choose $\mu_0$ as large as we need  in
Condition  \ref{condition of d}. Now we choose $0< c_0< c_1 <1$,
$\mu_0>4$ and $T$ satisfying the following condition: \bcd
\label{condition of c0c1c2T}
\begin{equation*}
\begin{cases}
\ds1) \;\;\; \mu_0 - 4c_1 -c_0 > 0,\\
\ns \ds 2)\;\;\; \frac{\mu_0}{(8c_1 +
c_0)}\sum_{i,j=1}^nb^{ij}d_{x_i}d_{x_j} > 4c_1^2T^2
> \sum_{i,j=1}^nb^{ij}d_{x_i}d_{x_j}.
\end{cases}
\end{equation*}
\ecd

\begin{remark}\label{rm2}
Since $\ds\sum_{i,j=1}^nb^{ij}d_{x_i}d_{x_j}
>0$, it is easy to see that one can always choose $\mu_0$ in Condition
\ref{condition of d} large enough so that Condition \ref{condition
of c0c1c2T} holds true. We put it here simply to emphasize the
relationship among $0< c_0< c_1 <1$, $\mu_0>4$ and $T$.
\end{remark}

In the sequel, we choose
\begin{eqnarray}\label{def gamma0}
\G_0 \= \Big\{ x\in \G :\,
\sum_{i,j=1}^nb^{ij}d_{x_i}(x)\nu^{j}(x) > 0
\Big\}.
\end{eqnarray}
Also, we assume that
\begin{equation} \label{aibi}
\begin{array}{ll}\ds
b_1 \in L_{\cal F}^{\infty}(0,T;L^{\infty}(G)),
\q b_2 \in L_{\cal
F}^{\infty}(0,T;L^{\infty}(G;\mathbb{R}^{n})),\\
\ns\ds
 b_3 \in L_{\cal F}^{\infty}(0,T;L^{n}(G)),\q\; b_4 \in L_{\cal
 F}^{\infty}(0,T;L^{\infty}(G)).
\end{array}
\end{equation}
In what follows, we use the notation:
\begin{eqnarray}
\cA\= |b_1|^2_{L_{\cal
F}^{\infty}(0,T;L^{\infty}(G))} +
|b_2|^2_{L_{\cal
F}^{\infty}(0,T;L^{\infty}(G;\mathbb{R}^{n}))}
+ |b_3|^2_{L_{\cal F}^{\infty}(0,T;L^{n}(G))} +
|b_4|^2_{L_{\cal
F}^{\infty}(0,T;L^{\infty}(G))} +1.
\end{eqnarray}

The main result of this paper can be stated as follows.

\begin{theorem}\label{uniqueness}
Let $b_i$ $(1 \leq i \leq 4)$ satisfy (\ref{aibi}), and let
Conditions \ref{condition of d} and \ref{condition of c0c1c2T} hold.
Assume that the solution $z\in H_T$ of (\ref{system1}) satisfies
that $\frac{\pa z}{\pa \nu}=0$ on $(0,T)\times \G_0$ and $z(T)=0$ in
$G$, $P$-a.s. Then $g=0$ in $Q$ and $z_0=z_1=0$ in $G$, $P$-a.s.
\end{theorem}

Several remarks are in order.

\begin{remark}
Similar to the inverse problem for (\ref{sy2}), and stimulated by
Theorem \ref{uniqueness}, it seems natural and reasonable to expect
a similar uniqueness result for the following equation
\begin{eqnarray}{\label{sy1}}
\left\{
\begin{array}{lll}\ds
\ds dz_{t} - \sum_{i,j=1}^n(b^{ij}z_{x_i})_{x_j}dt = \left(b_1 z_t +
b_2\cd\nabla z + b_3 z +f\right)dt + b_4z dB(t) & {\mbox { in }} Q,
 \\
\ns\ds  z = 0 & \mbox{ on } \Si, \\
\ns\ds  z(0) = z_0,\ z_{t}(0) = z_1 & \mbox{ in } G,
\end{array}
\right.
\end{eqnarray}
in which $z_0$, $z_1$ and $f$ are unknown and one expects to
determine them through the boundary observation $\left.\frac{\pa
z}{\pa \nu}\right|_{(0,T)\times \G_0}$ and the terminal measurement
$z(T)$. However the same conclusion as that in Theorem
\ref{uniqueness} does NOT hold true even for the deterministic wave
equation. Indeed,  we choose any $y\in C_0^\infty(Q)$  so that it
does not vanish in some proper nonempty subdomain of $Q$. Put
$f=u_{tt} - \D u$. Then, it is easy to see that $y$ solves the
following wave equation
 $$
\left\{
\begin{array}{ll}\ds
y_{tt} -\D y = f & \mbox{ in }Q,\\
\ns\ds y=0, &\mbox{ on } \Si,\\
\ns\ds y(0)=0,\ y_t(0)=0 &\mbox{ in }G.
\end{array}
\right.
 $$
One can show that $y(T)=0$ in $G$ and $\frac{\pa y}{\pa \nu}=0$ on
$\Si$. However, it is clear that $f$ does not vanish in $Q$. This
counterexample shows that the formulation of the stochastic inverse
problem may differ considerably from its deterministic counterpart.
\end{remark}

\begin{remark}\label{rm3}
From the computational point of view, it is quite interesting to
study the following stability problem (for the inverse stochastic
hyperbolic equation (\ref{system1})):  Is the map
 $$
 \left.\frac{\pa z}{\pa \nu}(z_0,z_1,g)\right|_{(0,T)\times \G_0}\times
 z(z_0,z_1,g)(T)\longrightarrow (z_0,z_1,g)
 $$
continuous in some suitable Hilbert spaces? Unfortunately, we are
not able to prove this stability result at this moment. Instead,
from the proof of Theorem \ref{uniqueness} (See Theorem
\ref{carleman th} in Section 4), it is easy to show the following
partial stability result, i.e., for any solution $z\in H_T$ of the
equation (\ref{system1}) satisfying $z(T)=0$ in $G$, $P$-a.s., it
holds that
 $$
 |(z_0, z_1)|_{L^2(\O,{\cal F}_0, P; H^1_0(G)
\t L^2(G))}+|\sqrt{T-t}g|_{ L^2_{\cal
 F}(0,T;L^2(G))}\le C\left|\frac{\pa z}{\pa \nu}\right|_{L^2_{\cal
 F}(0,T;L^2(\G_0))}.
 $$
Especially, if  $g$ is of the form (\ref{po1}) (with
$g_1(\cd,\cd)\in L^2_{\cal
 F}(0,T)\setminus \{0\}$ and $g_2(\cd)\in L^2(G)$), then the following
estimate holds
  $$
 |(z_0, z_1)|_{L^2(\O,{\cal F}_0, P; H^1_0(G)
\t L^2(G))}+|g_2|_{L^2(G)}\le C\left|\frac{\pa z}{\pa
\nu}\right|_{L^2_{\cal
 F}(0,T;L^2(\G_0))}.
 $$
\end{remark}

\begin{remark}\label{rm4}
The inverse problem considered in this work is a sort of inverse
source problems. It would be quite interesting to study the global
uniqueness and stability of inverse coefficient problems for
stochastic PDEs but this remains to be done, and it seems to be a
very difficult problem.
\end{remark}

\begin{remark}\label{rm5}
It is also interesting to study the same inverse problems but for
other stochastic PDEs, say the stochastic parabolic equation, the
stochastic Sch\"odinger equation, the stochastic plate equation and
so on. However, it seems that the technique developed in this paper
cannot be applied to these equations.
\end{remark}


\section{Some preliminaries }


In this section, we collect some preliminaries which will be used
later.

First, we show the following hidden regularity result for the
solution $z$ to the equation \eqref{system1} (This result means that
the observation of the normal derivative of $z$ makes sense, i.e.,
$|\frac{\partial z}{\partial \nu} |_{L^2_{\cal
 F}(0,T;L^2(\G_0))} < +\infty$).
\medskip
\begin{proposition}\label{hidden r}
Let $b_i$ $(1 \leq i \leq 4)$ satisfy (\ref{aibi}). Then, for any
solution of the equation (\ref{system1}), it holds that
\begin{equation}\label{hidden ine}
\Big|\frac{\partial z}{\partial \nu}\Big
|_{L^2_{\cal
 F}(0,T;L^2(\G))} \leq e^{C\cA} \left[|(z_0,z_1)|_{L^2(\O,{\cal F}_0, P;
H_0^1(G)\t L^2(G))} + |g|_{L^2_{\cal
 F}(0,T;L^2(G))}\right].
\end{equation}
\end{proposition}
\medskip
\begin{remark}
In \cite{Zhangxu3}, the author proved Proposition \ref{hidden r}
when $(b^{ij})_{1\leq i, j\leq n}$ is an identity matrix. The proof
of Proposition \ref{hidden r} for the general coefficient matrix
$(b^{ij})_{1\leq i, j\leq n}$ is similar, and therefore we give
below only a sketch of the proof.
\end{remark}

\medskip

{\it Proof of Proposition \ref{hidden r}}\,:
Since $\G\in C^2$, one can find a vector field
$h=(h^1,\cdots,h^n)\in
C^1(\mathbb{R}^n;\mathbb{R}^n)$ such that
$h=\nu$ on $\G$ (\see \cite{Lions0}). A direct
computation shows that
\begin{equation}\label{equality hidden1}
\begin{array}{ll}\ds
-\sum_{i=1}^n\Big[ 2(h\cd\n z)\sum_{j=1}^n b^{ij}z_{x_j} + h^i\Big(
z_t^2 - \sum_{i,j=1}^n
b^{ij}z_{x_i} z_{x_j} \Big) \Big]_{x_i}dt\\
\ns\ds = 2\Big\{ \Big[dz_t - \sum_{i,j=1}^n
(b^{ij}z_{x_i})_{x_j}\Big] h \cd \n z - d(z_t
h\cd \n z) - \sum_{i,j,k=1}^n b^{ij}z_{x_i}
z_{x_k}
h^k_{x_j}\Big\}dt\\
\ns\ds \qq -z_t^2 \div h dt + \sum_{i,j=1}^n
z_{x_j} z_{x_i} \div(b^{ij}h)dt.
\end{array}
\end{equation}
Integrating the identity \eqref{equality hidden1} in $Q$, taking
expectation in $\O$ and using integration by parts, we obtain the
inequality \eqref{hidden ine} immediately. \endpf

\medskip

Next, we recall the following known result.

\begin{lemma} {\rm (\cite{Zhangxu3})} \label{hyperbolic1}
Let $p^{ij} \in C^{1}((0,T)\t \mathbb{R}^n)$
satisfy
\begin{equation}\label{pij}
p^{ij} = p^{ji}, \qq i,j =
1,2,\cdots,n,
\end{equation}
$\ell ,\,f,\,\Psi \in C^2((0,T)\t\mathbb{R}^n)$. Assume that $u$ is
an $H^2_{loc}(\mathbb{R}^n)$-valued $\{\cF_t\}_{t\geq 0}$-adapted
process such that $u_t$ is an $L^2(\mathbb{R}^n)$-valued
semimartingale. Set $\theta = e^\ell $ and $v=\theta u$. Then, for
a.e. $x\in \mathbb{R}^n$ and P-a.s. $\o \in \O$,
\begin{eqnarray}\label{hyperbolic2}
\begin{array}
{ll} &\ds \theta \Big( -2\ell _t v_t + 2\sum_{i,j=1}^np^{ij}\ell
_{x_i} v_{x_j} + \Psi v
 \Big) \Big[ du_t -
\sum_{i,j=1}^n(p^{ij}u_{x_i})_{x_j} dt \Big]\\
\ns& \ds \q+\sum_{i,j=1}^n\Big[ \sum_{i',j'=1}^n\big(
2p^{ij}p^{i'j'}\ell _{x_{i'}}v_{x_i}v_{x_{j'}} - p^{ij}p^{i'j'}\ell
_{x_i}v_{x_{i'}}v_{x_{j'}}
\big) - 2p^{ij}\ell _t v_{x_i} v_t + p^{ij}\ell _{x_i} v_t^2 \\
\ns &\ds \qq\qq + \Psi p^{ij}v_{x_i} v - \Big( A\ell _{x_i} +
\frac{\Psi_{x_i}}{2}\Big)p^{ij}v^2 \Big]_j dt \\
\ns &\quad \ds  +d\Big[ \sum_{i,j=1}^np^{ij}\ell _t v_{x_i} v_{x_j}
- 2\sum_{i,j=1}^np^{ij}\ell _{x_i}v_{x_j}v_t + \ell _t v_t^2 - \Psi
v_t v + \Big( A\ell _t +
\frac{\Psi_t}{2}\Big)v^2 \Big]  \\
\ns  = & \ds \Big\{ \Big[ \ell _{tt} + \sum_{i,j=1}^n(p^{ij}\ell
_{x_i})_{j} - \Psi \Big]v_t^2 - 2\sum_{i,j=1}^n[(p^{ij}\ell
_{x_j})_t +
p^{ij}\ell _{tj}]v_{x_i}v_t\\
\ns & \quad \ds +\sum_{i,j=1}^n \Big[ (p^{ij}\ell _t)_t +
\sum_{i',j'=1}^n\Big(2p^{ij'}(p^{i'j}\ell
_{x_{i'}})_{j'}-(p^{ij}p^{i'j'}\ell _{x_{i'}})_{j'}\Big)
+ \Psi p^{ij} \Big]v_{x_i}v_{x_j}  \\
\ns & \quad \ds + Bv^2 + \Big( -2\ell _tv_t +
2\sum_{i,j=1}^np^{ij}\ell _{x_i}v_{x_j} + \Psi v \Big)^2\Big\} dt +
\theta^2\ell _t(du_t)^2,
\end{array}
\end{eqnarray}
where $(du_t)^2$ denotes the quadratic
variation process of $u_t$, $A$ and $B$ are
stated as follows:
\begin{eqnarray}\label{AB1}
\left\{
\begin{array}{lll}\ds
A\=(\ell _t^2 - \ell _{tt}) - \sum_{i,j=1}^n(p^{ij}\ell _{x_i}\ell
_{x_j} -p_{x_j}^{ij}\ell _{x_i} -
p^{ij}\ell _{x_i x_j})-\Psi, \\
\ns\ds B\=A\Psi + (A\ell _t)_t-\sum_{i,j=1}^n(Ap^{ij}\ell
_{x_i})_{x_j} + \frac{1}{2}\Big[ \Psi_{tt} -
\sum_{i,j=1}^n(p^{ij}\Psi_{x_i})_{x_j} \Big].
\end{array}
\right.
\end{eqnarray}
\end{lemma}

\section{Proof of the main result}

This section is devoted to proving  Theorem \ref{uniqueness}. As
mentioned before, we will prove Theorem \ref{uniqueness} by
establishing a new Carleman estimate for the equation
(\ref{system1}).

In the sequel, we choose
 $$\theta = e^\ell, \qq \ell  = \l\big[d(x) - c_1 ( t-T )^2\big],
 $$
where $\l>0$ is a parameter, $d(\cd)$ is the function given in
Condition \ref{condition of d}, and $c_1$ is the constant in
Condition \ref{condition of c0c1c2T}.

Our global Carleman estimate for \eqref{system1} is as follows.
\begin{theorem}\label{carleman th}
Let $b_i$ $(1 \leq i \leq 4)$ satisfy (\ref{aibi}), and let
Conditions \ref{condition of d} and \ref{condition of c0c1c2T} hold.
Then, there exists a constant $\tilde\l>0$ such that for any
$\l\geq\tilde \l$ and any solution $z\in H_T$ of the equation
(\ref{system1}) satisfying $z(T)=0$ in $G$, $P$-a.s., it holds that
\begin{equation}\label{carleman1}
\begin{array}{ll}\ds
\mathbb{E}\int_G\theta^2 ( \l|z_1|^2 + \l |\nabla z_0|^2 + \l^3
|z_0|^2)dx + \l\mathbb{E}\int_Q (T-t) \theta^2g^2dxdt\\
\ns\ds \leq C  \l \mathbb{E} \int_0^T\int_{\G_0}\theta^2\Big|
\frac{\pa z}{\pa\nu} \Big|^2 d\G dt.
\end{array}
\end{equation}
\end{theorem}

{\it Proof}\,: In what follows, we shall apply Lemma
\ref{hyperbolic1} to the equation (\ref{system1}) with
 $$u=z, \q p^{ij} = b^{ij}, \q \Psi
= \ell _{tt} + \sum_{i,j=1}^n(b^{ij}\ell _{x_i})_{x_j} - \l  c_0,
 $$
(Recall Condition \ref{condition of c0c1c2T} for the constant
$c_0$), and then estimate the terms in (\ref{hyperbolic2}) one by
one.

In the sequel, for $\l>0$, we use $O(\l^r)$ to denote a function of
order $\l^r$ for large $\l$. The proof is divided into three steps.

\ms

{\bf Step 1.} In this step, we analyze the terms which stand for the
``energy" of the solution of (\ref{system1}). As the Carleman
estimate for deterministic partial differential equation, the point
is to compute the order of $\l$ in the coefficients of $|v_t|^2$,
$|\nabla v|^2$ and $|v|^2$. Since the computation is very close to
that in the proof of \cite[Theorem 1.2.1]{Luqi2}, we give here only
a sketch.

First,  it is clear that the coefficient of $|v_t|^2$ reads:
\begin{equation}\label{coeffvt}
\ell _{tt} + \sum_{i,j=1}^n(b^{ij}\ell _{x_i})_{x_j} -\Psi = \l c_0
.
\end{equation}
Further, noting that $b^{ij}$ ($1 \leq i, j \leq n$) are independent
of $t$ and   $\ell _{t x_j}=\ell _{x_j t}=0$, we find that
\begin{equation}\label{coeffvtvi}
\sum_{i,j=1}^n(b^{ij}\ell _{x_j})_t + b^{ij}\ell _{t x_j}v_{x_i} v_t
= 0.
\end{equation}
Further, by Condition \ref{condition of d},  we see that
\begin{eqnarray}
\begin{array}{ll} & \ds\sum_{i,j=1}^n\Big\{ (b^{ij}\ell _t)_t + \sum_{i',j'=1}^n\Big[
2b^{ij'}(b^{i'j}\ell _{x_{i'}})_{x_{j'}} - (b^{ij}b^{i'j'}\ell
_{x_{i'}})_{x_{j'}} \Big] +
\Psi b^{ij} \Big\}v_{x_i} v_{x_j} \\
\ns & \ds \geq \l (\mu_0 -4c_1 - c_0)\sum_{i,j=1}^nb^{ij}v_{x_i}
v_{x_j}.
\end{array}
\end{eqnarray}
Further, in order to compute the coefficient $B$ of $|v|^2$,
recalling (\ref{AB1}), we find that
\begin{eqnarray}
&\,&A =  \l^2  \Big[ 4c_1^2(t - T)^2 - \sum_{i,j=1}^n b^{ij}d_{x_i}
d_{x_j} \Big] + O(\l).
\end{eqnarray}
Hence, by the definition of $B$ (in (\ref{AB1})), we conclude that
\begin{eqnarray}
\begin{array} {ll} \ds B  & \ds = (4c_1+c_0) \sum_{i,j=1}^n b^{ij}d_{x_i}d_{x_j} \l^3 +
\sum_{i,j}^n\sum_{i',j'=1}^n b^{ij} d_{x_i}(b^{i'j'}d_{x_{i'}}d_{x_{j'}})_{x_j}\l^3  \\
\ns &\ds\quad - 4(8c_1^3 + c_0c_1^2)(t-T)^2\l^3 + O(\l^2).
\end{array}
\end{eqnarray}
Recall the following estimate in \cite{Luqi2}:
\begin{eqnarray}
\mu_0\sum_{i,j=1}^nb^{ij}d_{x_i}d_{x_j}\leq
\sum_{i,j=1}^n\sum_{i',j'=1}^nb^{ij}d_{x_i}(b^{i'j'}d_{x_{i'}}d_{x_{j'}})_{x_j}.
\end{eqnarray}
Therefore, by Condition \ref{condition of c0c1c2T}, we obtain that
\begin{eqnarray}
 B \ds & \geq& (4c_1+c_0)\sum_{i,j=1}^nb^{ij}d_{x_i}d_{x_j}\l^3 + \mu_0
\sum_{i,j=1}^n b^{ij}d_{x_i}d_{x_j}\l^3\nonumber\\
\ns & & \ds  - 4(8c_1
+ c_0)c_1^2(t-T)^2\l^3 + O(\l^2) \\
\ns &=&\ds (4c_1+c_0)\sum_{i,j=1}^nb^{ij}d_{x_i}d_{x_j}\l^3 +
O(\l^2).\nonumber
\end{eqnarray}
Hence, there exists a $\l_0 > 0$ such that for any $\l \geq \l_0$,
it holds that
\begin{equation}\label{B ine}
Bv^2 \geq C\l^3 v^2.
\end{equation}

\ms

{\bf Step 2.} In this step, we analyze the terms corresponding to
$t=0$ and $t=T$. For the time $t=0$, we have
\begin{eqnarray}
\begin{array}{ll} \ds \sum_{i,j=1}^nb^{ij}\ell _tv_{x_i}v_{x_j} - 2
\sum_{i,j=1}^nb^{ij}\ell _{x_i}v_{x_j}v_t + \ell _tv_t^2 -\Psi v_t v
+ \Big( A\ell _t + \frac{1}{2}\Psi_t \Big)v^2
\\
\ns   =  \ds  2c_1 T \l\sum_{i,j=1}^nb^{ij}v_{x_i}v_{x_j} - 2\l
\sum_{i,j=1}^nb^{ij}d_{x_i}v_{x_j}v_t  - \l\Big(-2c_1 +
\sum_{i,j=1}^n(b^{ij}d_{x_i})_{x_j} -c_0\Big)v_t v\\
\ns  \q\ds + 2c_1 T \l v_t^2+ \Big[ 2c_1 T\Big(4c_1^2T^2 -
\sum_{i,j=1}^nb^{ij}d_{x_i}d_{x_j}\Big)\l^3 + O(\l^2) \Big]v^2
\\
\ns  \geq  \ds 2c_1 T \l\sum_{i,j=1}^nb^{ij}v_{x_i}v_{x_j} - \l
\Big( \sum_{i,j=1}^nb^{ij}d_{x_i}d_{x_j}
\Big)^{\frac{1}{2}}\sum_{i,j=1}^nb^{ij}v_{x_i}v_{x_j} - \l \Big(
\sum_{i,j=1}^nb^{ij}d_{x_i}d_{x_j} \Big)^{\frac{1}{2}}v_t^2  \\
\ns \q\ds+ 2c_1 T \l v_t^2 - v_t^2 + \Big[2c_1T\Big(4c_1^2T^2 -
\sum_{i,j=1}^nb^{ij}d_{x_i}d_{x_j}\Big)\l^3 + O(\l^2)\Big]v^2.
\end{array}
\end{eqnarray}
By Condition (\ref{condition of c0c1c2T}), it follows that
$$4c_1^2T^2 - \sum_{i,j=1}^nb^{ij}d_{x_i}d_{x_j} > 0$$  and that
$$2c_1T - \Big( \sum_{i,j=1}^nb^{ij}d_{x_i}d_{x_j}
\Big)^{\frac{1}{2}} > 0.$$
Hence there exists a $\l_1 > 0$ such that for any $\l \geq \l_1$ and
when $t=0$, it holds that
\begin{equation}\label{d0}
\begin{array}{ll}\ds
\sum_{i,j=1}^nb^{ij}\ell _tv_{x_i}v_{x_j} - 2
\sum_{i,j=1}^nb^{ij}\ell _{x_i}v_{x_j}v_t + \ell _tv_t^2 -\Psi v_t v
+ \Big( A\ell _t + \frac{1}{2}\Psi_t \Big)v^2
\\\ns\ds \geq C\Big[\l(v_t^2 + |\nabla v|^2) +
\l^3 v^2 \Big].
\end{array}
\end{equation}

On the other hand, since $\ell _t(T)=0$, for $t=T$, it holds that
\begin{eqnarray}\label{dT}
\begin{array}
{ll} \ds \sum_{i,j=1}^nb^{ij}\ell _tv_{x_i}v_{x_j} - 2
\sum_{i,j=1}^nb^{ij}\ell _{x_i}v_{x_j}v_t + \ell _tv_t^2 -\Psi v_t v
+ \Big(
A\ell _t + \frac{1}{2}\Psi_t \Big)v^2\\
\ns =  \ds - 2 \sum_{i,j=1}^nb^{ij}\ell _{x_i}v_{x_j}v_t -\Psi v_t
v.
\end{array}
\end{eqnarray}
Noting that $z(T)=0$ in $G$, $P$-a.s., we have
$v (T)= 0$ and $v_{x_j}(T)=0$ in $G$
($j=1,2,\cdots,n$), $P$-a.s. Thus, from the
equality \eqref{dT}, we end up with
\begin{eqnarray}\label{dT1}
\3n\3n\3n\left.\left\{\sum_{i,j=1}^nb^{ij}\ell _tv_{x_i}v_{x_j} - 2
\sum_{i,j=1}^nb^{ij}\ell _{x_i}v_{x_j}v_t + \ell _tv_t^2 -\Psi v_t v
+ \Big( A\ell _t + \frac{1}{2}\Psi_t
\Big)v^2\right\}\right|_{t=T}=0, \; P\mbox{-a.s.}
\end{eqnarray}

\ms

{\bf Step 3.} Integrating
(\ref{hyperbolic2}) in $Q$, taking expectation
in $\O$ and by the argument above, for
$\l\geq\max\{\l_0,\l_1\}$, we obtain that
\begin{eqnarray}\label{hyperbolic31}
\begin{array}
{ll} \ds\mathbb{E}\int_Q \theta\Big\{\Big( -2\ell _t v_t +
2\sum_{i,j=1}^nb^{ij}\ell _{x_i}v_{x_j} + \Psi v \Big) \Big[ dz_t -
\sum_{i,j=1}^n(b^{ij}z_{x_i})_{x_j}dt \Big] \Big\}dx
\\
\ns  \quad \ds + \l
\mathbb{E}\int_{\Si}\sum_{i,j=1}^n\sum_{i',j'=1}^n\Big(
2b^{ij}b^{i'j'}d_{x_{i'}}v_{x_i} v_{x_{j'}} - b^{ij}b^{i'j'}d_{x_i}
v_{x_{i'}}v_{x_{j'}} \Big)\nu_j d\Si\\
\ns   \geq  \ds C\Big\{\mathbb{E}\int_Q \Big[\theta^2\Big( \l z_t^2
+ \l |\nabla z|^2 + \l^3 z^2 \Big) + \Big(-2\ell _tv_t +
2\sum_{i,j=1}^nb^{ij}\ell _{x_i}v_{x_j} + \Psi
v\Big)^2\Big]dxdt  \\
\ns \q\ds + \mathbb{E} \int_G \theta^2\Big[\l(|\nabla z_0|^2 +
|z_1|^2) + \l^3|z_0|^2 \Big]dx  + \mathbb{E}\int_Q\theta^2 \ell _t
(dz_t)^2 \Big\}.
\end{array}
\end{eqnarray}
For the boundary term, noting that $z=0$ on $\Si$, it is easy to
show that
\begin{eqnarray}\label{hyperbolic32}
\begin{array}
{ll} \ds \mathbb{E}\int_{\Si}\sum_{i,j=1}^n\sum_{i',j'=1}^n\Big(
2b^{ij}b^{i'j'}d_{x_{i'}}v_{x_i} v_{x_{j'}} - b^{ij}b^{i'j'}d_{x_i}
v_{x_{i'}}v_{x_{j'}}
\Big)\nu_j d\Si \\
\ns  =  \ds \mathbb{E}\int_{\Si}\Big( \sum_{i,j=1}^nb^{ij}\nu_{x_i}
\nu_j \Big)\Big( \sum_{i',j'=1}^nb^{i'j'}d_{x_{i'}}\nu_{j'}
\Big)\Big|\frac{\pa v}{\pa \nu}\Big|^2d\Si.
\end{array}
\end{eqnarray}

From inequality (\ref{hyperbolic31}) and equality
(\ref{hyperbolic32}), we obtain that
\begin{eqnarray}\label{hyperbolic3}
\begin{array}
{ll} \ds \mathbb{E}\int_Q \theta\Big\{\Big( -2\ell _t v_t +
2\sum_{i,j=1}^nb^{ij}\ell _{x_i}v_{x_j} + \Psi v \Big)
\Big[ du_t - \sum_{i,j=1}^n(b^{ij}u_{x_i})_{x_j}dt \Big]\Big\}dx  \\
\ns  \quad \ds+ \l \mathbb{E}\int_{\Si}\Big(
\sum_{i,j=1}^nb^{ij}\nu_{x_i} \nu_j \Big)\Big(
\sum_{i',j'=1}^nb^{i'j'}d_{x_{i'}}\nu_{j'} \Big)\Big|\frac{\pa
v}{\pa
\nu}\Big|^2d\Si  \\
\ns  \geq \ds C\Big\{\mathbb{E}\int_Q \Big[\theta^2\Big( \l  z_t^2 +
\l |\nabla z|^2 + \l^3 z^2 \Big) + \Big(-2\ell _tv_t +
2\sum_{i,j=1}^nb^{ij}\ell _{x_i}v_{x_j} + \Psi
v\Big)^2\Big]dxdt  \\
\ns \quad \ds + \mathbb{E} \int_G \theta^2\Big[\l(|\nabla z_0|^2 +
|z_1|^2) + \l^3|z_0|^2 \Big]dx + \l\mathbb{E}\int_Q (T-t) \theta^2
(b_4z + g)^2 dxdt\Big\}.
\end{array}
\end{eqnarray}

By means of $$  (b_4 z + g)^2 \geq \frac{1}{2} g^2 - 2b_4^2 z^2,$$
we get
\begin{equation}\label{hyperbolic3.1}
\l\mathbb{E}\int_Q (T-t) \theta^2 (b_4z + g)^2
dxdt \geq \frac{1}{2}\l\mathbb{E}\int_Q (T-t)
\theta^2  g^2   dxdt - 2\l T\mathbb{E}\int_Q
\theta^2  b_4^2 z^2  dxdt.
\end{equation}
On the other hand, by equation (\ref{system1}), it is clear that
\begin{eqnarray}\label{hyperbolic4}
\begin{array}
{ll}
 \ds \mathbb{E}\int_Q \theta\Big\{\Big( -2\ell _t v_t +
2\sum_{i,j=1}^nb^{ij}\ell _{x_i}v_{x_j} + \Psi v \Big) \Big[ dz_t -
\sum_{i,j=1}^n(b^{ij}z_{x_i})_{x_j}dt \Big] \Big\}dx
 \\
\ns  \leq  \ds  \mathbb{E}\int_Q\Big(-2\ell _t v_t +
\sum_{i,j=1}^nb^{ij}\ell _{x_i} v_{x_j} + \Psi v \Big)^2dxdt
 + C\bigg\{|b_1|^2_{L^{\infty}_{\cF}(0,T;L^{\infty}(G))} \mathbb{E}\int_Q
\theta^2 z_t^2 dxdt
\\
\ns  \quad \ds +
\Big[|b_2|^2_{L^{\infty}_{\cF}(0,T;L^{\infty}(G,\mathbb{R}^n))}
+
|b_3|^2_{L^{\infty}_{\cF}(0,T;L^{n}(G))}\Big]\mathbb{E}\int_Q\theta^2
|\nabla z|^2dxdt
 \\
\ns  \quad \ds  +
\l^2|b_3|^2_{L^{\infty}_{\cF}(0,T;L^{n}(G))}\mathbb{E}\int_Q
\theta^2 z^2 dxdt   \bigg\}.
\end{array}
\end{eqnarray}

Finally, taking $\tilde\l = \max\big\{C\cA,
\l_0, \l_1\big\}$, combining (\ref{def
gamma0}), (\ref{hyperbolic3}),
(\ref{hyperbolic3.1}) and (\ref{hyperbolic4}),
for any $\l \geq \tilde \l$, we conclude the
desired estimate (\ref{carleman1}).\endpf

\ms

We are now in a position to prove Theorem \ref{uniqueness}.

\vspace{0.2cm}

{\it Proof of Theorem \ref{uniqueness}}\,: Since $\frac{\pa
z}{\pa\nu}=0$ on $\Si_0$, $P$-a.s., we know the right hand side of
inequality (\ref{carleman1}) is zero. Therefore, it follows that
\begin{equation}\label{invine1}
\mathbb{E}\int_G\theta^2 ( \l|z_1|^2 + \l
|\nabla z_0|^2 + \l^3 |z_0|^2)dx=0
\end{equation}
and that
\begin{equation}\label{invine2}
\mathbb{E}\int_Q (T-t) \theta^2g^2dxdt=0.
\end{equation}
From the equality \eqref{invine1}, we find $z_0 = z_1 =0$ in $G$,
$P$-a.s. By means of the equality \eqref{invine2}, we see $g=0$ in
$Q$, $P$-a.s.\endpf




\end{document}